\documentclass[aps,twocolumn,prb,showpacs,preprintnumbers,amsmath,amssymb,superscriptaddress]{revtex4}

\usepackage{graphicx}% Include figure files
\usepackage{dcolumn}% Align table columns on decimal point
\usepackage{bm}% bold math

\begin{document}

\title{Fluctuations, dissipation, and nonuniversal superfluid jumps in two-dimensional superconductors}

\author{R. W. Crane}
\affiliation{Department of Physics and Astronomy, University of
California, Los Angeles, CA 90095}

\author{N.P. Armitage }
\email{npa@pha.jhu.edu}

\affiliation{Department of Physics and Astronomy, University of
California, Los Angeles, CA 90095}

\affiliation{D\'{e}partement de Physique de la Mati\`{e}re
Condens\'{e}e, Universit\'{e} de Gen\`{e}ve, quai Ernest-Ansermet
24, CH1211 Gen\`{e}ve 4, Switzerland}

\affiliation{Department of Physics and Astronomy, The Johns
Hopkins University, Baltimore, MD 21218 }

\author{A. Johansson}
\affiliation{Department of Condensed Matter Physics, Weizmann
Institute of Science, Rehovot 76100, Israel}

\author{G. Sambandamurthy}\affiliation{Department of Condensed
Matter Physics, Weizmann Institute of Science, Rehovot 76100,
Israel}
\affiliation{SUNY - Buffalo, Department of Physics, 239
Fronczak Hall, Buffalo, NY 14260-1500}

\author{D. Shahar}
\affiliation{Department of Condensed Matter Physics, Weizmann
Institute of Science, Rehovot 76100, Israel}

\author{G. Gr\"{u}ner}
\affiliation{Department of Physics and Astronomy, University of
California, Los Angeles, CA 90095}

\date{\today}

\begin{abstract}

We report a comprehensive study of the complex AC conductivity of
thin effectively 2D amorphous superconducting InO$_x$ films at
zero applied field. Below a temperature scale $T_{c0}$ where the
superconducting order parameter amplitude becomes well defined,
there is a temperature where both the generalized superfluid
stiffness acquires a frequency dependence and the DC
mangetoresistance becomes linear in field.  We associate this with
a transition of the Kosterlitz-Thouless-Berezinskii (KTB) type. At
our measurement frequencies the superfluid stiffness at $T_{KTB}$
is found to be larger than the universal value. Although this may
be understood with a vortex dielectric constant of $\epsilon_v
\approx 1.9$ within the usual KTB theory, this is a relatively
large value and indicates that such a system may be out of the
domain of applicability of the low-fugacity (low vortex density)
KTB treatment. This opens up the possibility that at least some of
the discrepancy from a non-universal magnitude is intrinsic. Our
finite frequency measurements allow us access to a number of other
phenomena concerning the charge dynamics in superconducting thin
films, including an enhanced conductivity near the amplitude
fluctuation temperature $T_{c0}$ and a finite dissipation at low
temperature which appears to be a universal aspect of highly
disordered superconducting films.

\end{abstract}

% Superconductivity: Fluctuations; SC: Transport Properties; Optical Properties: Low-d films; OP: Microwaves

%Use showkeys class option if keyword

\maketitle

\section{Introduction}

Superconducting fluctuations have been an area of investigation
for many years \cite{Larkinreview,SkojTink}, yet very many
completely central issues even in conventional superconductors are
not understood \cite{Benfatto,SachdevKTB}.  Recently, interest in
superconducting fluctuation phenomena has been renewed and indeed
been thrust under the $klieg$ lights due to focus on the
high-temperature superconductors, where many believe them to play
a central role in the phenomenology of underdoped compounds
\cite{EmeryKivelson,Ong,Corson}.

In the usual view, as one approaches the nominal superconducting
transition temperature T$_{c}$ from above, thermal fluctuations in
the amplitude \cite{amplitudedisc} of the complex superconducting
order parameter (OP) $\Psi = \Delta e ^{i \phi}$ occur. If the
fluctuations are small, they can be expressed in terms of a
Gaussian approximation within the Ginzburg-Landau (GL) theory.  In
such a treatment it is envisioned that the free energy
fluctuations of the OP's amplitude inside a coherence volume
$\xi^3$ are of order $k_B T$. In the context of the microscopic
theory, this physics can be found as the Aslamazov-Larkin (AL)
contribution to the conductivity \cite{AslaLarkin,Tinkham}.

In a typical good 3D BCS superconductor such fluctuations occur
only vanishingly close to $T_{c}$.  However the fluctuation regime
can be significantly enhanced for a thin dirty short coherence
length ($\xi$) superconductor (thickness $d$) where the coherence
volume becomes $d \xi^2$.  In actuality true long range order of a
continuous OP cannot exist for $T>0$ in a 2D system, but the
expectation is that there is still a crossover temperature
T$_{c0}$ where the superconducting amplitude becomes relatively
well defined, although the phase still fluctuates.  It is believed
an actual phase transition does not occur until a possibly much
lower temperature where a transition to a topologically phase
ordered state can take place. Such a so-called
Kosterlitz-Thouless-Berezinskii (KTB) \cite{Berez,KToriginal}
transition can be written formally in a system with 2D $xy$ or
U(1) symmetry as a vortex-antivortex unbinding transition from a
state of matter with a generalized rigidity (the superfluid
stiffness) and power-law correlations to one with no rigidity and
only exponential correlations.  Above the transition, the
superconducting phase fluctuates via unbound vortex excitations.
It is the finite superfluid stiffness at low temperature that
allows the phase to assume a well-defined value. The superfluid
stiffness is proportional to the superfluid density and the
expectation is that the KTB transition is typified by a
discontinuous jump in this parameter, which in certain limits
takes on a universal value given by the transition temperature
itself $N = \frac{8 m }{\pi \hbar^2} k_B T_{KTB}$.  It is
relatively well established that a KTB transition with a universal
jump occurs in thin superfluid He4 layers \cite{KTBhelium},
although the situation in thin homogeneously disordered
superconductor films is far from clear.

For the case of superconductors, KTB physics has been mainly
investigated via linear and non-linear DC transport
\cite{Beasley,Fiory1983,Hebard1983,Hebard1985,Hsu1992,Kadin} where
various predictions exist for the temperature activated resistance
and power law of the IV characteristics.  It has been found that
the experimental situation is roughly consistent with theoretical
predictions, with, for instance, the $\alpha$ exponent of $V
\varpropto I ^\alpha$ a decreasing function of temperature until
dropping quickly to become $\sim 3$ at a temperature not so far
from the expected $T_{KTB}$.  Measurements of the frequency
dependent conductivity should, in principle, allow more specific
tests of theories for various fluctuation regimes and crossovers
between them due to the sensitivity of the probe to the time scale
of the fluctuations themselves.  Moreover, finite frequency
measurements lend themselves to the study of both real and
imaginary components of the charge response which gives more
information and allows a more precise comparison to theory.  In
relatively thick films, a few groups have shown via infrared and
microwave transmission in granular or low disorder lead and
aluminum\cite{LeadAC,AlAC1,AlAC2,AlAC3} that there was a region
around the bulk $T_{c}$ that had superconducting fluctuations
consistent with the dynamic AL amplitude fluctuation form. In high
disorder thin films, using much lower frequencies, a number of
groups have shown a dependence roughly consistent with the
predictions of the dynamic extension of the KTB formalism
\cite{Fiory1983,Hebard1980,Turneaure,Fiory,Yazdani}, although
certain discrepancies exist.

So although various aspects of the commonly accepted picture have
been reported, in-depth characterization of the finite frequency
dynamics of both the Gaussian and vortex regimes have not been
made, nor has a clear picture emerged as to how the system evolves
from one regime of dominant fluctuations to the other, before
ultimately entering the low temperature phase. Moreover, it has
never been shown whether or not these signatures of KTB physics
actually occur at a superfluid stiffness given by the universal
jump condition and by extension exactly what the nature of the low
temperature state is. In this paper we present a comprehensive
study of the AC conductivity of highly disordered InO$_x$ films at
GHz frequencies. We find that below a temperature $T_{c0}$ where
the superconducting amplitude becomes well defined, the films
exhibit a temperature region where the generalized frequency
dependent superfluid stiffness acquires a frequency dependence. We
observe that at essentially the same temperature the DC
magnetoresistance becomes linear in applied field.  This is the
generic behavior expected near a KTB transition, however in the
present case it is exhibited at a stiffness well above the
predicted universal value for the inferred $T_{KTB}$.  Although it
is possible to understand this within the usual KTB theory with a
vortex dielectric constant of $\epsilon_v \approx 1.9$, this
relatively large $\epsilon_v$ may indicate that the vortex
fugacity is too large in such a system for the conventional low
vortex density KTB treatment.  This opens up the possibility that
at least some of the non-universal value of the superfluid
stiffness at T$_{KTB}$ is intrinsic. At the lowest temperatures,
we continue to observe a finite dissipation, which we correlate to
an inhomogeneous superfluid density distribution. These results
are put in the context of the considerable existing literature.

\section{Experimental Details}

For these measurements, high purity (99.999 \%) In$_2$O$_3$ was
e-gun evaporated under high vacuum onto clean 1mm thick 19mm wide
sapphire discs.  Our synthesis methods are patterned off of the
work of Ref. \cite{Ovadyahu} where it was shown that amorphous
InO$_x$ can be reproducibly made by a combination of e-beam
evaporation of In$_2$O$_3$ and possibly annealing at low
temperatures. This is unlike films prepared via other methods
\cite{HebardNakahara,Zhu}. Essentially identical films have been
used in a large number of recent studies of the 2D
superconductor-insulator quantum phase transition
\cite{Steiner1,Steiner2,Murthy04,Murthy05,Murthy06,CranePRL,Gantmakher}.
Thin machined aluminum masks were used to pattern the films
creating a 200 $\AA$-thick 3mm-wide circular amorphous film
centered on the disc. Sample deposition was well controlled and
for a certain conditions samples can be made reproducibly
\cite{Murthy04,Steiner2}.  For structural characterization, we
co-deposit two more films along with the sample: (1) onto a TEM
grid for electron diffraction. (2) for AFM scans. We believe that
the films are morphologically homogeneous with no crystalline
inclusions or large scale morphological disorder for the following
reasons:

1. The TEM-diffraction patterns are diffuse rings with no
diffraction spots, suggesting amorphous films with no crystalline
inclusions.

2. The AFM images show continuous films with no voids or cracks
and in fact are completely featureless down to a scale of a few nm
(the resolution of the AFM).

3. The R vs T curves when investigating the 2D
superconductor-insulator transition \cite{Murthy04,CranePRL} in
these films are smooth with no re-entrant behavior that is the
hallmark of gross inhomogeneity.

AC conductivity was measured in a novel cryomagnetic resonant
microwave cavity system.  The cavity diameter was optimized for
performance in the 22 GHz ($\hbar \omega / k_B = 1.06$ K) TE011
mode.  A number of other discrete frequencies from 9 to 106 GHz
were accessible by insertion of an additional sapphire puck (for
the low frequencies) or use of a very short "pan"-shaped cavity
(for greater than 100 GHz).  Our highest operating frequency of
106 GHz corresponds in temperature units (using $\hbar \omega /
k_B$) to 5.09 K, but as we will point out below, using the BCS
relation $2 \Delta / k_B T_{c} = 3.53$ and the temperature scale
where the amplitude becomes well defined (2.28 K) the threshold
for above gap excitation is 158 GHz.  The lack of sample heating
was ensured by operating in regime where the response was
independent of input power. Relations between the resonances'
frequency shift $\Delta \omega$ and change in quality factor
$\Delta (1/Q)$ upon sample introduction to the complex
conductivity are obtained by a cavity perturbation technique
\cite{KotzlerBrandt1,KotzlerBrandt3,Redbook,Peligrad} (see
\cite{Waldron} for a very thorough treatment). This standard
experimental technique is based on the adiabatic modification of
the electromagnetic fields in a cavity that arises from the
introduction of a small sample to the interior of the cavity. Our
data are analyzed in the depolarization regime in which the fields
in the cavity penetrate the entire volume of the sample.  It has
been shown that for extremely thin films, only in-plane AC
electric fields or out-of-plane AC magnetic fields at the sample
position can affect an appreciable change in a cavity's resonance
characteristics \cite{Peligrad2}. Samples were placed along the
cavity's central axis, where due to symmetry consideration and
depending on the particular TE mode being used, if the electric
field is in-plane then there is a zero out-of-plane magnetic field
and vice versa.

Modes with both these field configurations were exploited in our
setup and their analysis differed. The formulas used for analysis
will be stated here for completeness in terms of the complex
frequency $\hat{\omega} \equiv \omega_{0} - i\omega_{0}/2Q$. For
the case of an in-plane electric field, the complex frequency
shift is
    \begin{eqnarray}
    \frac{\delta\hat{\omega}}{\omega} =
    -\gamma\frac{4 \pi \sigma^*/\omega}{1+n 4 \pi \sigma^*/\omega}
    \end{eqnarray}

where $\sigma^*$ is the complex conjugate of the conductivity, $n$
is the depolarization factor, $\gamma =
\gamma_{0}\frac{V_{sample}}{V_{cavity}}$ is a "filling factor",
and $\gamma_{0}$ is a constant that depends on the particular mode
used. In the present case the sample is so thin that $n$ is very
small ($3 \times 10^{-6}$) and so there is only very weak mixing
between components of the complex frequency shift into the complex
conductivity, so that the real part of the conductivity is largely
proportional to the change in $\omega_{0}/2Q$ and the imaginary
part of the conductivity is largely proportional to the shift in
$\omega_{0}$.

For the case of the sample being in a perpendicular AC magnetic
field the results are obtained following \cite{KotzlerBrandt3}
    \begin{eqnarray}
    \frac{\delta \tilde{\omega}}{\omega} = - \gamma \chi(\omega)
    \end{eqnarray}
where $\gamma$ is again the ``filling factor'' and $\chi(\omega)$
is the susceptibility which includes not only the response from
localized electric and magnetic dipoles, but also the response
from induced AC currents.  This is relevant for our analysis in a
magnetic field anti-node since the response from the local
electric dipole moments will be zero (electric field node), the
local magnetic moments are zero (non-magnetic material), and so
the cavity perturbation is due to the induced currents in the
superconductor which has been fully penetrated by the external
rf-field.  Brandt \cite{KotzlerBrandt1,BrandtTheoryPRB} has worked
out the susceptibility for a thin superconducting disk in a
perpendicular DC field due to in-plane AC currents from a
superimposed AC field and found
\begin{eqnarray}
\chi(\omega) =
-\sum_{n=1}^{N}\frac{C_{n}/\Lambda_{n}^{2}}{\Omega^{-1}+\Lambda_{n}^{-1}}
= -\frac{1}{\gamma}\frac{\delta\tilde{\omega}}{\omega}
\label{eq:BrandtSum}
\end{eqnarray}
where the quantity we wish to extract, namely the dynamic
conductivity $\sigma(\omega)$ is contained in the normalized
frequency $\Omega = i \omega \sigma(\omega)\cdot\frac{\mu_{0}r
d}{2\pi}$ where $r$ is the radius and $d$ the thickness of the
sample.  The coefficients $C_{n}$ and $\Lambda_{n}$ are given in
Ref. \cite{KotzlerBrandt3}, where it is found that to well within
experimental accuracy, the series can be represented by a small
number of terms.  We use 30 elements - as in previous studies -
but in reality many less are actually needed. With these relations
in hand, the complex conductivity can be extracted from the
complex frequency shift. Here also, due to the thinness of the
samples, there is only weak mixing between real and imaginary
frequencies into the components of the complex conductivity.

Although theoretical values exist for $\gamma$, in practice the
conversions to complex conductivity were made by adjusting the
free parameters for the two schemes ($\gamma$ and $n$ for the
E-field case and $\gamma$ for the B-field case) so that the AC
resistance matched DC data at temperatures well above the
occurrence of superconductivity and using the expectation that the
superfluid stiffness (defined below), was frequency
\textit{independent} at H=0 and low temperature.  With a unique
sample dependent depolarization factor and $\gamma$'s that depend
on the particular mode being used a unique set of normalization
factors could be found.

DC resistance was measured on co-deposited samples in a two-probe
configuration by low frequency AC lock-in techniques using
excitation currents of approximately $10 nA$. The probe's lead
resistances, which have a negligible temperature dependence in the
displayed temperature range, were well characterized and have been
subtracted from the displayed data.

\section{Results}

We are interested in understanding the fluctuation behavior of the
complex order parameter $\Delta e ^{i \phi}$ as the temperature is
lowered towards the superconducting state. A finite frequency
probe has access to this information because the complex
electrodynamic response of a superconductor at experimental
frequencies $\omega_{exp}$ below the gap $2 \Delta$ can be
approximated as $\sigma_{1} = (\pi n_{s} e^{2}/2 m)\delta(\omega)
+ n_{n} e^{2} \tau_{n}/m$ and $\sigma_{2} = n_{s} e^{2}/m \omega$,
where $n_{s}$ ($n_{n}$) is the density of superconducting (normal)
electrons, $\tau_{n}$ is the relaxation time for normal electrons,
and $e$ and $m$ are the electron charge and mass respectively.  We
note that, in fact, in a highly disordered superconductor the
contribution from normal electrons is very small as the carrier
lifetime is very short.  In particular, the normal electron
contribution to $\sigma_2$ is negligible when $\omega \tau \ll 1$
as certainly the case here in our highly disordered material.  As
will be shown below, also importantly for our study, the extremely
short normal state lifetime causes even the above gap contribution
to $\sigma_2$ for $\hbar \omega > 2\Delta$ to be below detectable
levels for our experimental frequencies. This is important and
means that our $\sigma_2$ is still almost purely due to the
superfluid response even at elevated temperatures when the
superconducting gap begins to close.

For a fluctuating superconductor one can define $\sigma_{2} =
n_{s}(\omega) e^{2}/m \omega$ where the fluctuation effects are
captured by a frequency-dependent superfluid density
$n_{s}(\omega)$ \cite{CranePRL,KTdynamics}.  A definition as such
- or in terms of an equivalent frequency dependent superfluid
stiffness (defined below) - has been the usual treatment within
the finite frequency KTB theory \cite{KTdynamics}, where AC
measurements allow one to probe the system on shorter length
scales and reveal superconducting fluctuations even above
$T_{KTB}$.

\begin{figure}[htb]
    \includegraphics[width=8.2cm,angle=0]{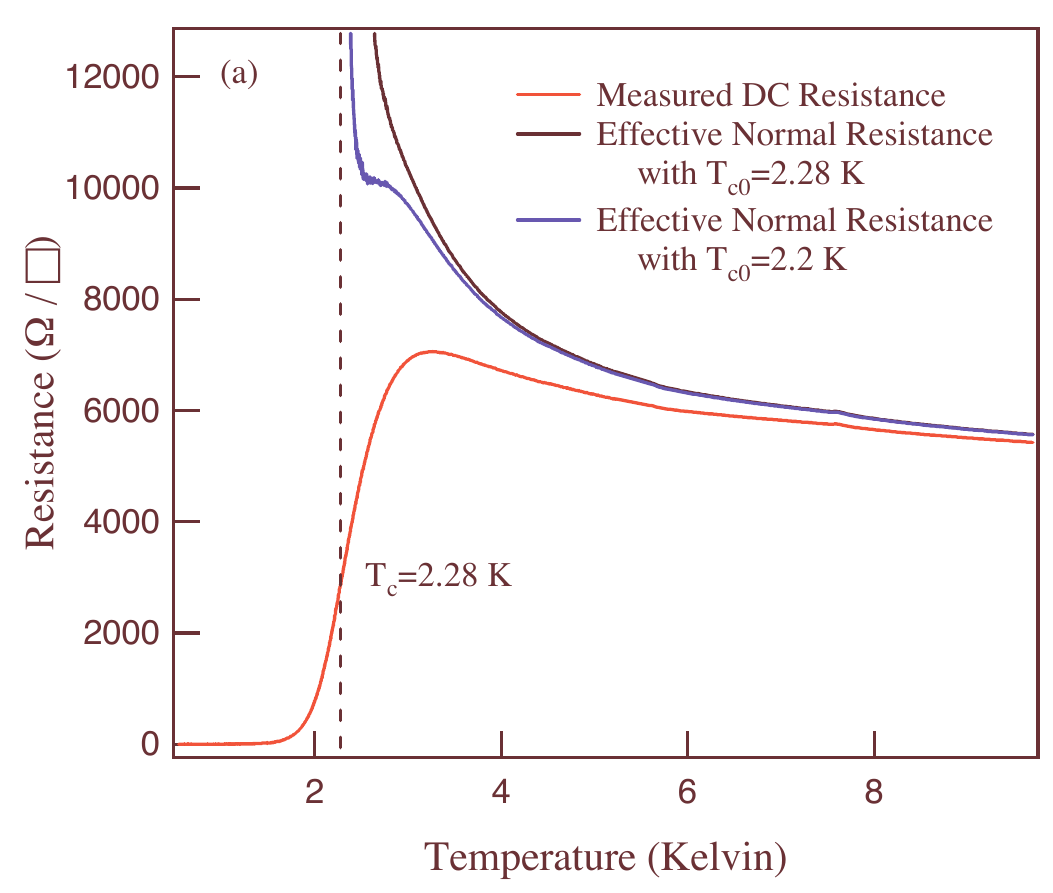}
    \caption{(color) DC
    Sheet Resistance.  (a) Temperature dependence showing the
    broad resistive transition resulting from fluctuations. The
    normal state resistance curves are generated using the
    procedure described in the text. T$_{co}$ is believed to be 2.28 K.
    \label{fig:DC}}
    \end{figure}

We begin by examining the DC data, as shown in Fig. \ref{fig:DC}.
We observe a broad region over which the superconducting
transition occurs.  The contribution of Gaussian amplitude
fluctuations can be obtained by fitting to the Aslamazov-Larkin DC
form.  Using the procedure of Gantmakher \cite{Gantmakher}, a
lower bound on $T_{co}$ can be estimated as the lowest temperature
that does not cause an inflection point in the extracted effective
normal state resistance $R_N(T)$ as defined by the full expression
for the Aslamazov-Larkin fluctuation resistivity.

    \begin{eqnarray}
    R_{meas} = \frac{1}{\sigma_N d + \sigma_{2D}^{AL} d} = \frac{1} { 1/R_{N} + \frac{ e^{2}  }{16 \hbar}
    \frac{ T_{c0}  }{(T-T_{c0})}} \label{eq:ALFluctuationConductivity}
    \end{eqnarray}

Within this analysis $2.28 K$ is the best lower bound on $T_{c0}$
and represents the temperature scale below which the
superconducting amplitude is relatively well defined. Lower values
of  $T_{c0}$ produce a kink in the extracted resistivity, where
for instance a $T_{c0}$ of $2.2K$ is clearly too low as seen in
Fig. 1a.

    %DCdata.pxp  ---->  July2006.pxp

As mentioned above, the imaginary conductivity $\sigma_{2}$ is
proportional to the superfluid density, which then sets the scale
for the superfluid stiffness - the energy scale to introduce phase
slips in the superconducting order parameter.  We extract this
generalized superfluid stiffness $T_{\theta}$ (in degrees Kelvin)
from the conductivity via the relation $\sigma_2 = \sigma_Q
\frac{k_B T_{\theta}}{\hbar \omega}$ (as done similarly recently
\cite{Corson}), where $\sigma_Q = \frac{4e^2}{hd} $ is the quantum
of conductance for Cooper pairs divided by the sample thickness.
We emphasize that in our notation $T_{\theta}$ \textit{is not a
temperature per se}, but is an energy scale expressed in
temperature units.  Our superfluid stiffness is essentially
kinetic inductance expressed in temperature units.  We could
express this quantity in any one of a number of equivalents units
\cite{Lobb} including superfluid density. We prefer this
parametrization as it is unnecessary to make an estimate for
quantities such as the Cooper pair mass. Moreover with these units
the universal jump condition for the KTB transition is of the
particularly simple form $4 T_{KTB} = T_{\theta}$. The
fluctuations of the superconducting state can be captured by the
frequency dependence of the stiffness, which would otherwise be
frequency independent.

In the absence of fluctuations, we expect an accurate description
of the superfluid stiffness would be given by the dirty limit
mean-field BCS theory, which predicts the onset of
superconductivity at a temperature $T_{c}$, with a temperature
dependence of $\sigma_{2}(T)$ given by \cite{Tinkham,Sheahen}

    \begin{eqnarray}
    \sigma_{2}(T) = \frac{\pi \Delta(T)}{\hbar\omega}
    \tanh\left[\frac{\Delta(T)}{2
    k_{B}T}\right] \label{eq:BCSdirty}
    \end{eqnarray}
where $\Delta(T) = \Delta(0)
\sqrt{\cos\left[\frac{\pi}{2}\left(\frac{T}{T_{c}}\right)^{2}
\right]}$ is the temperature dependent gap function and $\Delta(0)
= 1.76 k_{B}T_{c}$ \cite{Sheahen1966}.

In Fig. 2 we show the temperature dependence of the superfluid
stiffness $T_{\theta}$ measured at various frequencies alongside
the prediction of equation \ref{eq:BCSdirty}. The mean field curve
has been generated by specifying the conductivity in the normal
state $\sigma_{n}$ and then varying $T_{c}$ to obtain the best fit
to the data at the lowest temperature.  We extract a transition
temperature scale of 3.47 $K$ which we associate with a
microscopic scale $T_{\mu}$ where the superconducting transition
$would$ occur in the absence of fluctuations.  This very large
temperature scale is over 1.6 times the temperature $T_{c0}$ where
the amplitude becomes well defined and over 3 times the scale of
the onset of phase coherence (defined below) and shows the
paramount role that superconducting fluctuations play in such a
material.  We also see that the superfluid stiffness curves
progressively approach the mean field curve at higher frequencies.
However, because of the enhanced role of fluctuations all of the
curves approach zero at a temperature below $T_{\mu}$.

% ForPeter_v2.pxp
    \begin{figure}
    \includegraphics[width=8.4cm,angle=0]{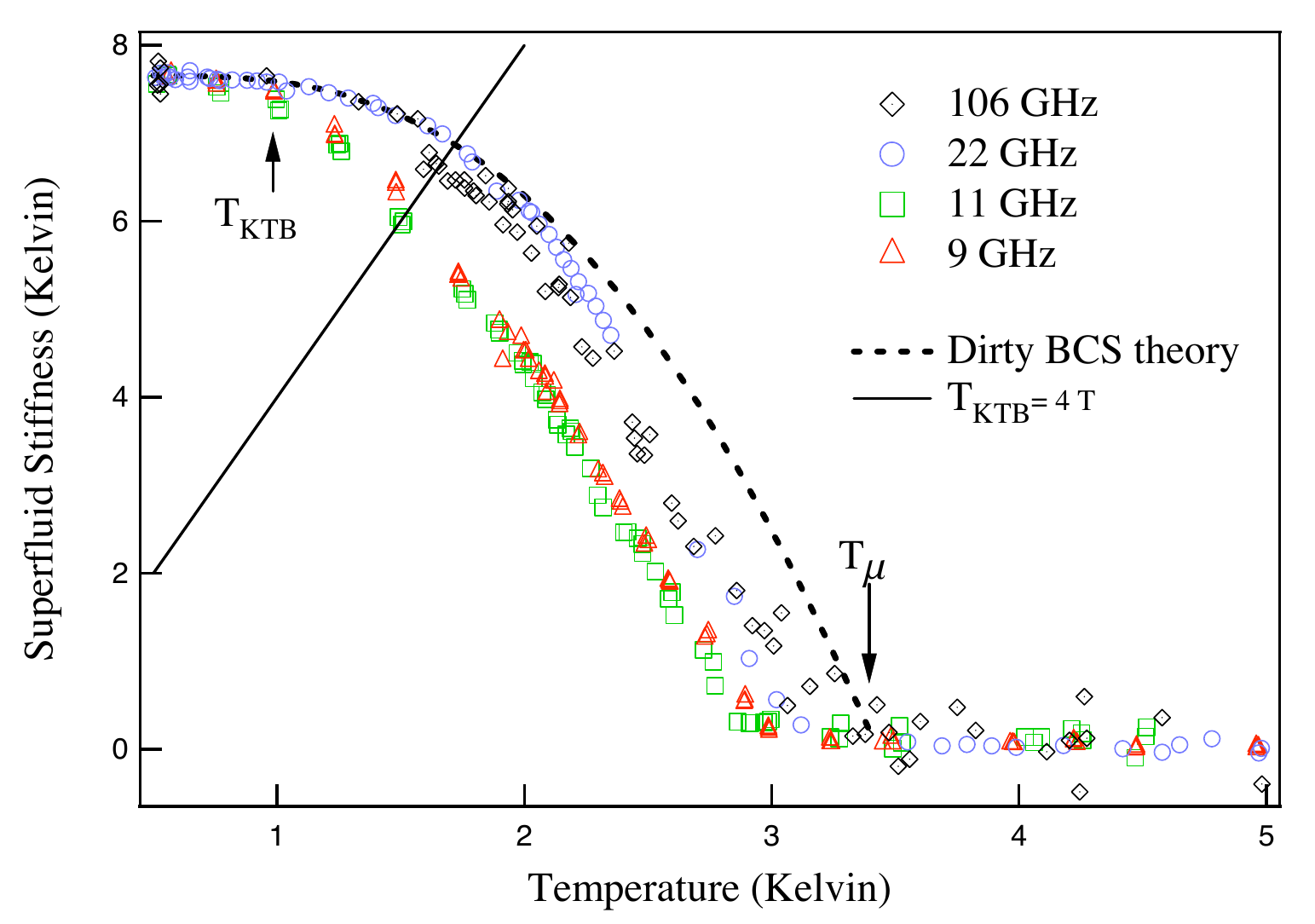}
    \caption{(color) Temperature dependence of the superfluid stiffness
    $T_{\theta}$.  The prediction of the dirty BCS model is shown
    as a dashed line which goes to zero at $T_{\mu}$.  The lower frequency data
    show a significant deviation from the mean field behavior.  The temperature where $T_{\theta}$ acquires
    a frequency dependence can be identified with $T_{KTB}$. \label{fig:dirtyBCS}}
    \end{figure}

The experimental curves in Fig. \ref{fig:dirtyBCS} show a broad
temperature region characterized by a gentle roll off of the
superfluid stiffness with increasing temperature.  We observe a
distinct temperature where the superfluid stiffness acquires a
frequency dependence. Within the standard theory, such an
occurrence is indicative of the approach to a KTB transition. In
this model, the \textit{zero-frequency} superfluid stiffness is
renormalized discontinuously to zero at a temperature $T_{KTB}$
set by the superfluid stiffness itself at this temperature.  Above
$T_{KTB}$ the system still appears superconducting on short length
scales set by the separation between thermally generated free
vortices and therefore at finite frequencies we expect the
superfluid stiffness to approach zero \textit{continuously}.  As
$T_{KTB}$ is the temperature where vortices proliferate, in at
least moderate fugacity (a quantity related to the vortex core
potential and defined below) superconductors, the temperature
where the superfluid stiffness curves measured at different
frequencies deviate from each other can be identified with the
approach to $T_{KTB}$.  As mentioned above, with our units of
superfluid stiffness the predicted transition temperature is $
T_{KTB} = T_{\theta}/4 $, which is shown as a solid line in the
figure.  We point out here that although a frequency dependence as
such is seen in Fig. \ref{fig:dirtyBCS}, the stiffness where
$T_{\theta}$ acquires its frequency dependence is well above the
stiffness predicted to be critical value.  This is presumably due
to the finite measurement frequency. We will expand on this point
below.

    %DCdata.pxp  ---->  July2006.pxp
    \begin{figure}[htb]
    \includegraphics[width=7.9cm,angle=0]{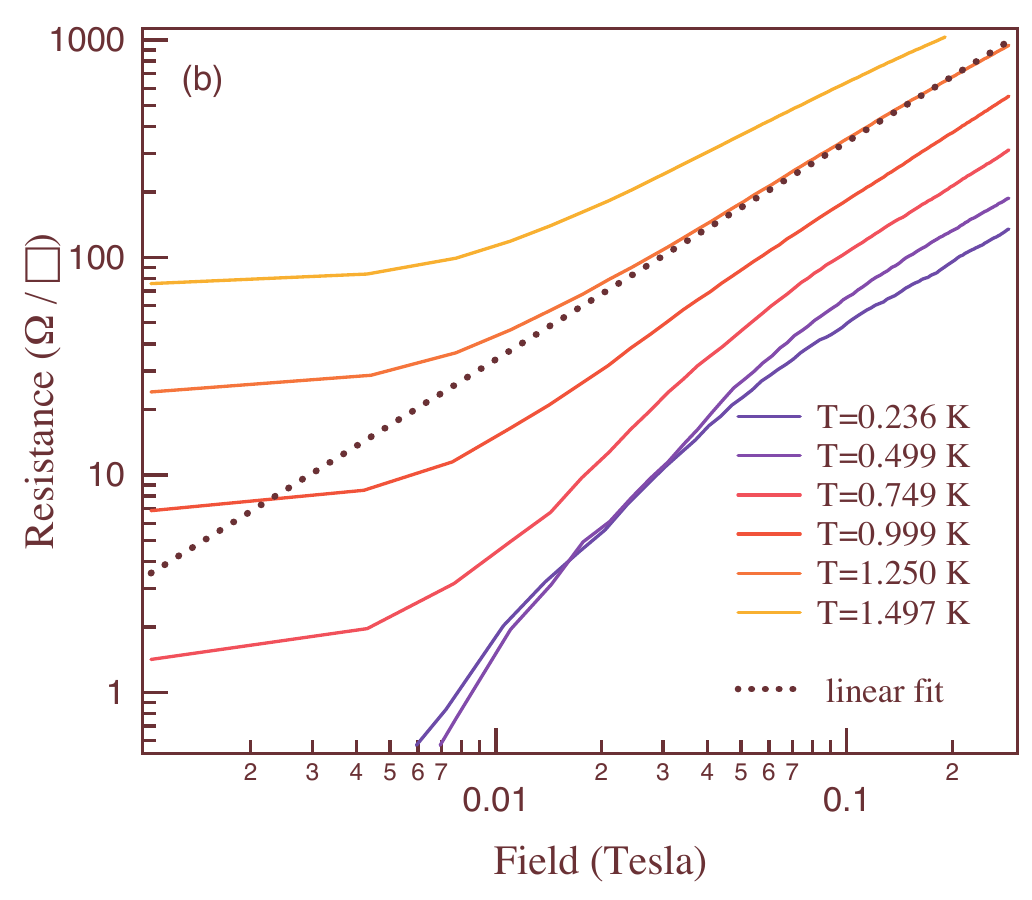}
     \caption{(color) The DC magnetoresistance isotherms showing a power law
    dependence on the applied field for temperature, T = 0.236,
    0.499, 0.749, 0.999, 1.250, and 1.497 Kelvin.  The dashed
    line is a linear fit.
    \label{fig:MR}}
    \end{figure}

Using the BCS relation $2 \Delta / k_B T_{c0} = 3.53$ and the
measured $T_{c0} = 2.16$ K, all operating frequencies are below
the BCS expectation of 158 GHz for above threshold gap excitation.
Although at the lowest temperatures our highest frequency is well
below the superconducting gap, in principle, there can be a normal
electron contribution to $\sigma_2$ from both thermally excited
quasiparticles at $T \neq 0$ as well as from above gap excitations
when $\hbar \omega > 2 \Delta$ on the approach to T$_c$.  In
practice, however these give a neglible contribution to the
response in a highly disordered superconductor.  We can give a
rough estimate of this contribution by a very approximate
calculation. Mangetoresistance measurements on similar samples
have shown that the coherence length is approximately 6 nm
\cite{Steiner2}.  As such highly disordered superconductors have
their coherence length set by the carrier mean free path, we can
take this value to be of the order of the normal electron
scattering length. With a reasonable estimate for the Fermi
velocity (c/200), this gives the very large effective scattering
rate ($1/\tau$) of approximately 200 - 300 THz.  This is
reasonable based on the high disorder and large normal state
resistivity of our sample.  An estimate based on the normal state
Drude conductivity and using a free electron mass and carrier
density estimated from Hall measurements on similar samples
($10^{21}/$cm$^3$) \cite{ShaharOvadyahu} gives a magnitude of the
same order.

Using this scattering rate, the Drude relations, our definition of
$T_{\theta}$, and its measured low temperature value (7.66 K) we
can estimate the maximum possible $contamination$ contribution to
the measured stiffness from "normal" electrons if all the spectral
weight in the superconducting delta function became subject to
normal state scattering as

    \begin{eqnarray}
T_{\theta ,cont} = T_{\theta}[T\rightarrow0] \frac{(\omega
\tau)^2}{1 + (\omega \tau)^2 }
    \label{eq:Tthetacont}
    \end{eqnarray}

At even our highest frequency, this represent a total contribution
of less than $2 \times 10^{-6}$ K, which is far smaller than our
sensitivity.  We note that even if we overestimated that effective
scattering rate by even a factor of 100 the contamination
contribution would only be 0.02 K which is still well below our
sensitivity. Hence, the contribution from ``normal'' electrons
give a completely insignificant contribution to $\sigma_2$ and
hence $T_{\theta}$.

As an aside we will note that an order of magnitude estimate of
the effective superconducting electron density can be found via
the definitions given above using the free electron mass and the
$T \rightarrow 0$ limit of $T_{\theta} = 7.66$ K gives an
approximate value of $3 \times 10^{17}/cm^{3}$ which can be
compared to the Hall effect derived number of $10^{21}/$cm$^3$
\cite{ShaharOvadyahu}. The effects of disorder are significant in
reducing the number of charges that participate in pairing.

Fig. \ref{fig:MR} shows the low field magnetoresistance data.  We
notice in the isotherms of magnetoresistance there is a range of
intermediate temperatures above the low field saturation regime
where the resistance changes from concave down to concave up with
magnetic field.  The condition $R(T) \propto B$ for low fields has
previously been associated with the
Kosterlitz-Thouless-Berezinskii temperature $T_{KTB}$
\cite{linearB} based on the Minnhagen criterion
\cite{Minnhagen1981} and was applied \cite{Hebard1990} to similar
samples of InO$_{x}$.

Power law fits to the low field data (well above the saturation
regime) of the magnetoresistance ($R \propto B^{\alpha}$) curves
gives a temperature dependent exponent in this range.  An
interpolation of the data gives a temperature of $\approx 1.20 K$
where $\alpha = 1$. Interestingly, this temperature is to within
the experimental uncertainty of where the superfluid stiffness
acquires a frequency dependence in Fig. \ref{fig:dirtyBCS}.  On
the basis of the above, we associate a temperature in the range of
$\approx 1.15 K$ with a transition of the KTB type involving
vortex unbinding and proliferation at a non-universal value of the
superfluid stiffness.

We also note that the zero-field resistance is finite just below
our inferred $T_{KTB}$. Such an observation is not surprising
within the usual KTB theory with our non-neglible excitation
current ($10nA$) and the possibility of small residual magnetic
fields, as the critical current and field of a 2D superconductor
are zero. Finite resistance below the inferred $T_{KTB}$ has been
observed before \cite{Hsu1992,Kadin}, but is not a universal
observation \cite{Hebard1980}.

Within the simplest conception of the KTB theory the transition
occurs at a value of the superfluid stiffness given by the above
universal jump condition, where the superfluid density is driven
to zero at T$_{KTB}$ by a cascade of vortex proliferation.  As
noted above, our frequency dependence is acquired in T$_{\theta}$
at a stiffness that is in excess of the universal prediction.  It
is important to realize that this prediction is for the zero
frequency stiffness (longest length scales) and higher frequency
probes can reveal larger values that don't take into account long
wavelength renormalizations.  Hence, transitions at non-universal
values of the bare (shortest length scales) superfluid stiffness
can occur where a large superfluid stiffness is renormalized
downward even below T$_{KTB}$ by thermally excited vortices if the
vortex potential for their activation is low enough. Our finite
frequency measurements allows us to probe the system on
intermediate length scales that are between these two limits.
Within the usual KTB theory, the transition will occur at a
universal value of the $renormalized$ superfluid stiffness where
the renormalization factor is given by a vortex dielectric
constant $1/\epsilon_v = 1 - 2 \pi y_0$ which incorporates the
screening effects of intervening thermally excited
vortex-antivortex pairs where $y_0 = e^{-\mu_c/k_B T}$ is the
so-called vortex fugacity (related to the thermal excitation
probability) and $\mu_c$ is the vortex core potential. In order
that the frequency dependence is observed at a T$_\theta$ of the
universal value at T$_{KTB}$, one must be specifically in the low
vortex density (low fugacity) limit. It is not clear that thin
film superconductors are generically in this limit.

We can estimate $\epsilon_v$ if we associate our extracted mean
field superfluid stiffness as the unrenormalized bare superfluid
value and take as the fully renormalized number the KTB prediction
at the temperature at which the frequency dependence in the
superfluid stiffness is acquired.  With this procedure we find a
$\epsilon_v$ of 1.9. This can be compared to a value of 1.3 for
both simulations of the 2D $xy$ model \cite{MinnhagenNylen} as
well as experiments on thin He4 films \cite{KTBhelium,Agnolet}.
According to the above fugacity relations this gives a core
potential $\mu_c$ of $2.97 K$. This core potential is a sum of
core energy and entropy terms $\mu_c = E_c + k_B T$ln$N_0$ where
$E_c$ is the core energy and $N_0$ represents the number of
statistically independent sites that two vortices can occupy
within a coherence area and is of the order of $\frac{1}{2 \pi}$.
Using this estimate we get a core energy of $0.85 K$, which is of
the expected magnitude\cite{disclaimer}. An understanding of our
results is not impossible within the usual KTB theory, however
this large $\epsilon_v$ is indicative of a relatively large
fugacity. As the usual KTB relations are derived in the limit of
low fugacity and hence low vortex density our results may mean
superconducting films as such are out of the domain of
applicability of the low fugacity theory.

A number of workers have raised the possibility that if the
fundamental assumption of low fugacity is violated, intrinsically
different physics may result, which may also be consistent with
our observations.  For instance, Minnhagen and coworkers have
postulated an expanded set of KTB renormalization equations valid
at larger fugacities and shown that out of the low fugacity limit
and above a critical value of $\epsilon_v \approx 1.74$ the
transition may exhibit different behavior, including a
non-universal jump of the renormalized superfluid stiffness and
perhaps even a first order transition \cite{Minnhagen,Jonsson}.
Such considerations may be relevant in our case with our
$\epsilon_v \approx 1.9$.  Additionally it has been proposed
\cite{GabayKapitulnik,Zhang} that at large fugacities vortices may
not form bound vortex-antivortex pairs below T$_{KTB}$ and
instead, their density is high enough that they crystalize into a
vortex ionic crystal. Superconducting phase coherence is disrupted
by the melting of this crystal, as opposed to the unbinding of
vortex-antivortex pairs.  Some evidence for such lattice formation
and subsequent melting exists in numerical simulations
\cite{Teitel}.

Irrespective of its origin, we associate this temperature
determined by both magnetoresistance and AC conductivity to be the
temperature scale where superconducting phase coherence is
destroyed.  Our analysis above has been made possible by the fact
that our highest measurement frequencies are large enough to be
close to the high frequency limit (as shown by the negligible
frequency dependence between 22 GHz and 106 GHz curves and their
relative closeness to the mean field curve) and thereby we can
measure the unrenormalized superfluid stiffness.

\begin{figure}
     \includegraphics[width=7.9cm,angle=0]{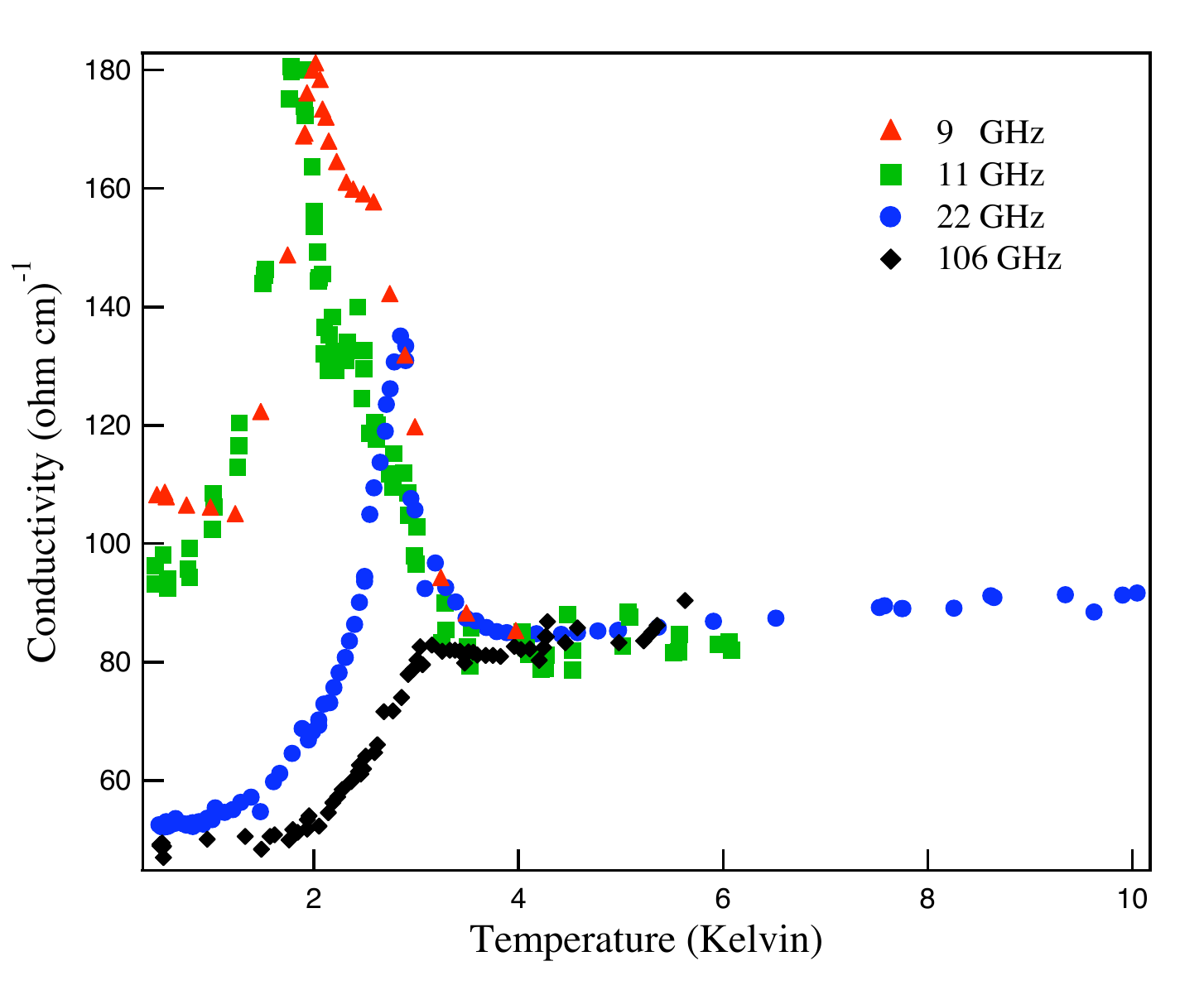}
    \caption{(color) Temperature dependence of $\sigma_{1}$ at 9, 11, 22
and 106 GHz. Note the shift in the peak to higher temperatures and
the decrease in its height with increasing frequency.
\label{fig:sigma1}}
    \end{figure}

Having established the temperature scales of the various
fluctuation phenomena in the problem we now discuss the
dissipative response of the real part of the conductivity
$\sigma_{1}$ (Fig. \ref{fig:sigma1}), focusing our attention on
the peak as well as the low temperature absorption.  We interpret
this dissipation peak in $\sigma_{1}$ as arising from a partially
coherent superfluid, as was conjectured previously in Ref.
\cite{Corson}. We observe that this peak, found in the temperature
region of amplitude fluctuations, shifts to higher temperatures
and its amplitude decreases as the frequency increases. Such an
effect is not predicted within the finite frequency Gaussian AL
theory where the peak in $\sigma_1$ is always at T$_c$
\cite{SkojTink,Tinkham}.

This behavior may be understood by comparing the length scale set
by the fluctuations and those set by the frequency of the
measurement probe.  At very high temperatures, well above the
fluctuation regime, there are no superconducting fluctuations. As
the temperature is lowered the characteristic length scale for
these fluctuations increases.  Higher frequencies probe the system
on shorter length scales, and therefore are able to measure the
onset of fluctuations at higher temperatures.  We would also like
to understand why there is a difference in the height of the
different dissipation peaks at different frequencies in Fig.
\ref{fig:sigma1}.  This may be understood heuristically by
returning to the ``dirty BCS theory'' mean field curve in Fig.
\ref{fig:dirtyBCS}, and noticing that as the temperature decreases
the superfluid stiffness is a rapidly increasing function of
temperature in this range. For a given measurement frequency, the
dissipation peak occurs when the timescale of the measurement
probe matches the timescale of the superconducting fluctuations,
and the peak height is proportional to the superfluid at that
temperature. Since this occurs at higher temperature for the
higher frequency probes, and the superfluid stiffness is a strong
function of temperature in this range, the peak height is expected
to decrease with increasing frequency.

\begin{figure}
    \includegraphics[width=8.5cm,angle=0]{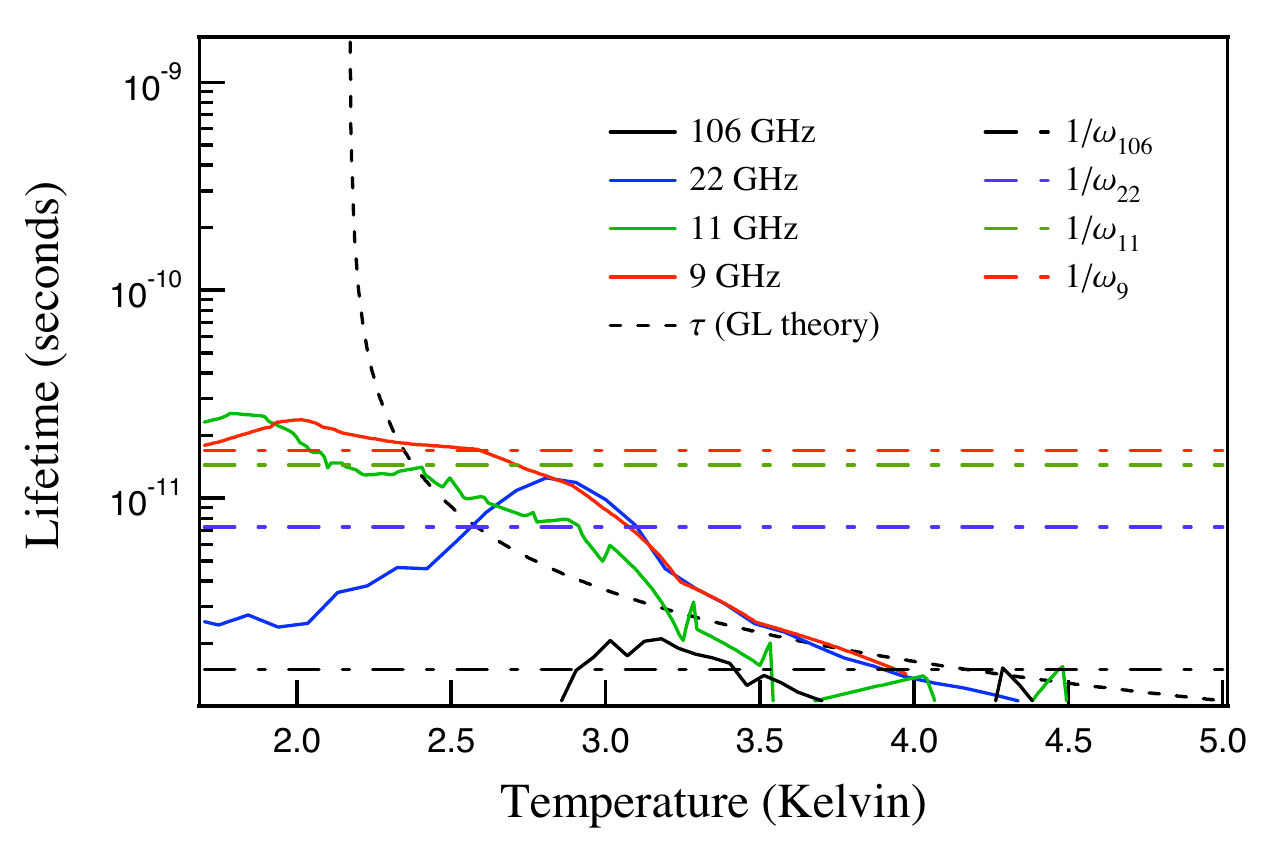}
     \caption{(color) $\tau$ vs temperature at various given frequencies extracted via the simple
    relaxation model in equation  \ref{eq:Relaxation}.  Also shown
    is the Ginzburg-Landau result using $T_{c0} = 2.28 K$.  The
    dashed horizontal lines mark the timescales set by the various probing frequencies,
    $1/\omega$.  Notice how the $\tau$ levels off at lower
    temperatures indicating the freezing out of amplitude
    fluctuations.
    \label{fig:tau}} \end{figure}

We can examine this further using a simple relaxation model
\cite{Waldram} which attempts to incorporate the observed
frequency dependence of $\sigma_{1}(T)$ in a simple way

    \begin{eqnarray}
    \tilde{\sigma}_{meas}(\omega,T) = \sigma_{N}^{DC}(T) +
    \frac{\sigma_{2D}^{AL}(T)}{1+i\omega\tau}
    \label{eq:Relaxation}
    \end{eqnarray}

where $\tilde{\sigma}_{meas}(\omega,T)$ is the measured finite
frequency conductivity, $\sigma_{N}^{DC}(T)$ is the normal DC
conductivity (fit using a $1/T$ temperature dependent
extrapolation \cite{Waldram,WaldramCites}) combined with the DC
Aslamazov-Larkin fluctuation conductivity from equation
\ref{eq:ALFluctuationConductivity} which appears to be the
substantial contribution to the resistivity in this temperature
region. We can use this to parameterize the data in terms of
$\tau$, which can then be compared with the Ginzburg-Landau result
$\tau_{GL} = \pi\hbar/8 k_{B} (T-T_{c0})$. This is shown in figure
\ref{fig:tau}.

The basic assumption of this approach is that there is a
contribution to $\sigma_{1}(\omega,T)$ in this temperature region
which - at zero frequency and temperatures above $T_{c0}$ - is
described by the Gaussian approximation given by the DC
Aslamazov-Larkin formula in equation
\ref{eq:ALFluctuationConductivity}. We then expect that the
$T_{c0}$ extracted from the DC data should appropriately capture
the underlying physics.  To account for the frequency dependence,
we assume a relaxation model in which the evanescent pairs
contribute to the conductivity for a time $\tau$ before decaying
back into normal electrons.  As shown in Fig. \ref{fig:tau}, such
a modelling qualitatively accounts for the frequency dependence as
evinced by the relative collapse of the $\tau$'s extracted at
different frequencies in Fig. 4. Importantly, we also see that
this treatment gives $\tau$ a plateau near $T_{c0}$ indicating a
slowing down of the fluctuation dynamics that we are sensitive to.
If, as we believe, the dissipation in this temperature range is
largely sensitive to the fluctuations of the superconducting
amplitude then the plateau is consistent with our interpretation
that there is a temperature T$_{c0}$ where the amplitude
fluctuations become frozen out. Finally, it is important to note
the temperature where the measured $\tau$ equals the inverse of
the measurement frequency $1/\omega$ occurs at approximately the
same temperature where we find a peak in the dissipative
($\sigma_{1}$) response in Fig. \ref{fig:sigma1}. This supports
the notion that the peak in $\sigma_{1}(T)$ arises when the
bandwidth of the probing frequency matches the lifetime of the
fluctuations.

Finally, we would like to point out the finite dissipation
($\sigma_{1}$) in Fig. \ref{fig:sigma1} in the limit $T
\rightarrow 0$, which is not accounted for by the BCS derived
theories of Mattis-Bardeen \cite{Tinkham,MattisBardeen} nor Leplae
\cite{Leplae}, which predict that dissipation disappears for
$\hbar \omega < 2 \Delta$ as $T \rightarrow 0$.  Anomalous
absorption as such has been seen previously in other highly
disordered superconducting films \cite{Karecki} as well as in
Bi-based high-temperature superconductors \cite{Corson2} which may
show nanoscale inhomogeneity \cite{PanDavis}.  The dissipation in
the case of the Bi-based high-temperature superconductors was
interpreted as a consequence of an inhomogeneous superfluid
density distribution.  Such a situation is certainly a possibility
here in our highly disordered films.  We reiterate, however, that
our amorphous films are themselves morphologically homogeneous
down to the lowest measured length scales (nanometers) as detailed
above.  It would be interesting to perform an scanning tunnelling
microscopy study to make a detailed comparison with the cuprate
superconductors.  That any inhomogeneity could arise out of films
which are essentially homogeneous on all measurable length scales
is the interesting and remarkable aspect. It is perhaps indicative
of an exponential sensitivity to the disorder which inevitably
exists at the shortest length scales (the materials are
amorphous). This extreme sensitivity to disorder and dynamically
created inhomogeneity is then interesting in its own right as it
is an intrinsic part of the problem \cite{Ghosal1,Ghosal2}.
Similar conclusion about dynamically created inhomogeneity have
been arrived at previously via DC transport based probes in
similar films \cite{Ovadyahu}.

We note that such an interpretation need not change our assumption
of a frequency independent superfluid stiffness at $H=0$ and $T
\rightarrow 0$. These other measurements that have seen low T
dissipation (for instance Ref. \cite{Corson2}) measure the
superfluid stiffness's absolute value directly and find that it is
frequency independent at low T. Additionally, we will point out
that explicit theoretical models that treat such a scenario
(within an effective medium granular scenario; for instance Ref.
\cite{Stroud}) with intermixed superconducting and metallic grains
finds that the $\sigma_2$ still has its characteristic $1/\omega$
dependence (and hence the superfluid stiffness in frequency
independent) for globally superconducting systems.  This derives
from the delta function in $\sigma_1$ at $\omega = 0$ which
Kramers-Kronig constrains the frequency dependence of $\sigma_2$.

Finally, we point out that as an alternative path to describing
this dissipation, it has been proposed that dissipative phenomena
may be generic in the vicinity of quantum critical points
\cite{Mason}. In this regard, our system is not very far in
parameter space to the 2D disorder-tuned superconductor-insulator
quantum phase transition \cite{Murthy04} and may be effected by
this physics.

\section{Conclusion}

In this work we have established various temperature scales
associated with the crossover between Gaussian and KTB-vortex
fluctuations.  We have been able to independently establish the
temperature of a KTB-like vortex unbinding transition by AC
techniques which allowed us to determine where the superfluid
stiffness acquired a frequency dependence, as well as by DC
measurements using the Minnhagen criterion.  The magnitude of the
superfluid stiffness disagrees with the prediction for the
universal superfluid jump criterion at $T_{KTB}$.  Although this
can be understood by invoking a relatively large vortex dielectric
constant $\epsilon_v \approx 1.9$, which can renormalize the
stiffness from a bare value to the universal value, it is also
indicative of a high fugacity, which may invalidate the
low-fugacity assumptions under which the KTB theory is derived.

We have also shown how the peak in the dissipative piece of the
complex response arises from superconducting fluctuations near
$T_{c0}$. In this model, the peak arises when the timescale of the
measurement matches the lifetime of the fluctuations and we have
given a simple model that causes a collapse in the relaxation time
measured at different frequencies. We give a simple picture that
describes this behavior and which shows how the dynamics slow upon
passing below the amplitude temperature scale.  Finally, we have
discussed how the dissipative piece of the complex conductivity
can be understood as arising from an inhomogeneous superfluid.

\section{Acknowledgements}

The authors would like to thank L. Benfatto, E.H. Brandt, and D.
Peligrad for useful correspondences, V. Zaretskey for help with
the instrumentation development and K. Holczer for experimental
support at a crucial time in this project. Research at UCLA was
supported by the NSF (DMR-0454540). Research at the Weizmann
Institute was supported by the ISF, the Koshland Fund and the
Minerva Foundation.


\begin{references}
\bibitem{amplitudedisc}  As is common in the literature, in this
paper we make a distinction between regimes of amplitude
fluctuations and regimes of phase fluctuations.  It is important
to keep in mind that, in fact, because the phase becomes
ill-defined as the amplitude vanishes, amplitude fluctuations
manifestly include a fluctuating phase.
\bibitem{disclaimer} Note using a small $N_0$ of 0.05 as was done in Ref. \cite{Kadin} gives
unphysical results for $E_c$.
\end{references}
\end{document}